\documentclass[12pt]{article}
\usepackage{amsbsy,amssymb}

\newcommand{\be}{\begin{eqnarray}}
\newcommand{\ee}{\end{eqnarray}}
\newcommand{\ben}{\begin{eqnarray*}}
\newcommand{\een}{\end{eqnarray*}}
\newcommand{\la}{\langle}
\newcommand{\ra}{\rangle}
\newcommand{\ba}{\mbox{\boldmath$\alpha$}}

\title{
KEPLER PROBLEM IN DIRAC THEORY FOR \\ A PARTICLE WITH POSITION-DEPENDENT
MASS}
\author{I. O. Vakarchuk\\
{\small Ivan Franko National University of Lviv}\\
{\small 12 Drahomanov Street, Lviv UA--79005, Ukraine}\\
{\small E-mail: chair@ktf.franko.lviv.ua }}
\date{ }
\begin{document}
\maketitle

\begin{abstract}
Exact solution of Dirac equation for a particle whose potential
energy and mass are inversely proportional to the distance from the force centre has been found.
The bound states exist provided the length scale $a$ which appears in the expression
for the mass is smaller than the classical electron radius $e^2/mc^2$. Furthermore,
bound states also exist for negative values of $a$ even in the absence of the Coulomb interaction.
Quasirelativistic expansion of the energy has been carried out, and a modified expression for the fine
structure of energy levels
has been obtained. The problem of kinetic energy operator in the Schr\"odinger equation is discussed for the case
of position-dependent mass. In particular, we have found that for highly excited states the mutual ordering of
the inverse mass and momentum operator in the non-relativistic theory is not important.

Key words:  effective mass, fine spectrum structure, radiation
corrections.
\end{abstract}
\vspace{0.5cm}

\noindent
%
%
\newpage
\section{Introduction}

The concept of the effective mass in theoretical physics is quite efficient
because it allows to reduce a many-body problem to a single particle one, without the loss of
the main contributions into the mechanism of the formation of various physical phenomena from the inter-particle
interactions. We can exemplify this statement by specific problems from the semiconductor physics,
superfluid $^4$He theories, problems of nanophysics as well as by a number of other
problems from the condensed matter physics \cite{Geller93,Serra97,Barranco97,Arias94,bastard}.
At the same time, the suggested approach raises the issue of mutual ordering of momentum operator in  the
Schr\"odinger equation
and the inverse effective mass in the kinetic energy operator, which is the momentum function of the particle
coordinate \cite{vonroos,levy,cavalcante,dutra,dutra1,dekar}.
This problem seems to disappear if one proceeds from the Dirac equation, but the
transition to the non-relativistic case is far from being simple as it might look at the first glance \cite{cavalcante}.

In this paper we suggest a solution of the Kepler problem (i.e., a study of the particle's movement
in the Coulomb potential) in the Dirac theory for a particle with the given effective mass $m^{*}$
dependent on coordinate  ${\bf r}$. We assume this dependence to be relevant at
distances of the particle from the force centre smaller than the Compton length $\lambda=\hbar/mc$,
where $m$ is the mass $m^{*}$ at $r\to\infty$.
For such distances the notion of the particle coordinate is lost as attempts to localize
the particle in the space domain with linear dimensions $\sim \lambda$ lead to the creation of new particles.
In this connection, it seems possible to try and take effectively into account
the processes of the interaction of the particles with the force centre at the subatomic scales
through a coordinate-dependent mass.

In the general case, we assume that, e.g., for an electron the value of $m^{*}$
can be presented in the form of a multipole expansion
\be
m^{*}=m\left(1+{a\over r}+{{\bf br}\over r^3}+\ldots\right),
\label{mul}
\ee
where $a$, ${\bf b}$, $\ldots$ are constants formed by the mechanism of the particle
interaction with the vacuum fluctuations in the presence of the force centre;
later on we consider these constants to be the given initial parameters of the problem.
The processes of the interaction at the subatomic scales within the quantum field theory lead, in particular,
to the deformation of the Coulomb interaction in the atom when the distance between the electron and the nucleus is small.
In other words, it is quite possible to consider the electron charge as a function of the particle coordinate.
We can therefore speak about the effective consideration, in the Dirac theory, of the radiation effects which are
due to the renormalization of the electron mass and charge; that is why we can make an attempt to
account, within such a phenomenological approach, for the observable superfine structure of the atom energy spectrum.

The above-mentioned arguments can be disregarded. Then, one treats the Dirac equation formally as a
mathematical problem in which the mass $m^{*}$ is dependent on the radius-vector ${\bf r}$.

Below, we provide an exact solution of the Kepler problem in the Dirac theory
for the case when only the first two terms are taken into account in the
expansion for the effective mass (\ref{mul}).

\section{Initial equations}
\setcounter{equation}{0}

We start from the Dirac equation (in familiar notation), i.e.,
\be
\left[\left(\hat{\mbox{\boldmath$\alpha$}}\hat{\bf p}\right)c+m^{*}c^2\hat{\beta}+U\right]\psi=E\psi,
\label{2.1}
\ee
where $\hat{\ba},\hat{\beta}$ are the Dirac matrices, the particle effective mass is
\be
m^{*} = m\left(1+{a\over r}\right),
\label{2.2}
\ee
and the energy of the Coulomb interaction is
\be
U=-{e^2\over r},
\label{2.3}
\ee
where $e$ is the particle charge.

In the following we calculate the energy spectrum $E$. To do this,
we introduce the function $\bar \psi$ so that
\be
\psi=\left[\left(\hat{\mbox{\boldmath$\alpha$}}\hat
{\bf p}\right)c+m^{*}c^2\hat{\beta}+(E-U)\right]\bar\psi,
\label{2.4}
\ee
that is, we apply the substitution used also in \cite{Green, Vakarchuk}.
For the new function $\bar \psi$ from the Dirac equation (\ref{2.1}) we find
\be
\left\{\left[\left(\hat{\mbox{\boldmath$\alpha$}}\hat
{\bf p}\right)c+m^{*}c^2\hat{\beta}\right]^2-c\left[\left(\hat{\mbox{\boldmath$\alpha$}}\hat
{\bf p}\right)U-U\left(\hat{\mbox{\boldmath$\alpha$}}\hat
{\bf p}\right)\right] -(E-U)^2\right\}\bar \psi=0.
\label{2.5}
\ee

It is easy to see that the commutator
\ben
\left[\hat{\mbox{\boldmath$\alpha$}}\hat
{\bf p}, U\right]=-i\hbar(\mbox{\boldmath{$\sigma$}}{\bf
n})\hat\beta' \, {dU\over dr},
\een
and
\be
\left[\left(\hat{\mbox{\boldmath$\alpha$}}\hat {\bf
p}\right)c+m^{*}c^2\beta\right]^2=\hat{\bf
p}^2c^2+m^{*2}c^4-i\hbar c^3(\mbox{\boldmath{$\sigma$}}{\bf
n})\hat\beta''\, {dm^{*}\over dr}, \label{2.6} \ee where
$\mbox{\boldmath{$\sigma$}}=(\hat\sigma_x,\hat\sigma_y,\hat\sigma_z)$
are the Pauli matrices,
\be
\hat\beta'=
\left(
\begin{array}{cl}
0&I\\
I&0
\end{array}
\right),
\ \ \ \ \
\hat\beta''=
\left(\begin{array}{ll}
0&-I\\
I&\,\,\,0
\end{array}
\right)
\ee
are $4\times4$ matrices, ${\bf n}={\bf r}/r$ is a unit vector.

Our initial equation (\ref{2.5}) now takes the following form
\be
\Bigg\{
\hat{\bf p}^2c^2+m^{*2}c^4-i\hbar c^3(\mbox{\boldmath{$\sigma$}}{\bf
n})\hat\beta''\, {dm^{*}\over dr}+i\hbar c(\mbox{\boldmath{$\sigma$}}{\bf
n})\hat\beta'\, {dU\over dr}-(E-U)^2
\Bigg\}\bar\psi=0.
\ee

Using explicit expresions for $m^{*}$ and $U$ (see Eqs.~(\ref{2.2})
(\ref{2.3})), we obtain, after simple calculations,
\be
&&\Bigg\{
{\hat{\bf p}^2\over 2m}-\left({E\over mc^2}e^2-mc^2a\right){1\over r}+{1\over 2mr^2}
\left[{i\hbar\over  c}(\mbox{\boldmath{$\sigma$}}{\bf n})
(mc^2a\hat\beta''+e^2\hat\beta')+m^2c^2 a^2-{e^4\over
c^2}\right]\Bigg\}\bar \psi\nonumber\\
&&=\left({E^2-m^2c^4\over 2mc^2}\right)\bar\psi.
\label{yavn}
\ee
We note that our equation has the form of the Schr\"odinger equation for a particle moving
in the Coulomb potential with an addant to the centrifugal energy.

\section{The radial equation}
\setcounter{equation}{0}

We now pass to spherical coordinates in Eq.~(\ref{yavn}):
\be
&&\Bigg\{
-{\hbar^2\over 2m}{1\over r}{d^2\over dr^2}\,r+{1\over 2mr^2}
\left[\hat{\bf L}^2+{i\hbar\over c}(\mbox{\boldmath{$\sigma$}}{\bf n})
(mc^2a\hat \beta''+e^2\hat\beta')+m^2c^2a^2-{e^4\over c^2}\right]
\nonumber\\
&&-\left({Ee^2\over mc^2}-mc^2a\right){1\over r}\Bigg\}\bar\psi={E^2-m^2c^4\over
2mc^2}\bar\psi,
\label{sfer1}
\ee
where $\hat{\bf L}$ is the angular momentum  operator.

As the operator in square brackets depends solely on the angles, the variables in Eq.~(\ref{sfer1}) can be separated.
We further make use of the fact that
\ben
\hat{\bf L}^2=(\mbox{\boldmath{$\sigma$}}\hat{\bf L})[(\mbox{\boldmath{$\sigma$}}\hat{\bf
L})+\hbar],
\een
and introduce the following operator
\be
\hat\Lambda=-[(\mbox{\boldmath{$\sigma$}}\hat{\bf L})+\hbar]+{i\over
c} (\mbox{\boldmath{$\sigma$}}{\bf n})(mc^2a\hat\beta''+e^2\hat
\beta'),
\label{lambd2}
\ee
with
\ben
\hat\Lambda^2=[(\mbox{\boldmath{$\sigma$}}\hat{\bf
L})+\hbar]^2+m^2c^2a^2-{e^4/c^2}.
\een
It is now not difficult to show that the operator in square brackets in Eq.~(\ref{sfer1}) equals
$\hat\Lambda(\hat\Lambda+\hbar)$. Then, we rewrite Eq.~(\ref{sfer1})
as follows
\be
&&\Bigg\{
-{\hbar^2\over 2m}{1\over r}{d^2\over dr^2}r+{\hat\Lambda(\hat\Lambda+\hbar)\over 2mr^2}
-\left({E\over mc^2}e^2-mc^2a\right){1\over r}\Bigg\}\bar\psi={E^2-m^2c^4\over
2mc^2}\,\bar\psi.
\label{sfer}
\ee

As $(\hat{\mbox{\boldmath{$\sigma$}}}\hat{\bf L})=(\hat {\bf J}^2-\hat {\bf L}^2
-\hat{\bf S}^2)/\hbar$, where the total angular momentum $\hat {\bf J}=\hat{\bf L}+\hat{\bf S},$
and the spin operator $\hat{\bf S}=\hbar\hat{\mbox{\boldmath{$\sigma$}}}/2$, it is easy to see that
the operator
\ben
\hat\Lambda^2=\left({\hat{\bf J}^2-\hat{\bf L}^2-\hat{\bf S}^2\over
\hbar}+\hbar\right)^2+(mca)^2-\left({e^2\over c}\right)^2.
\een

It then follows that the eigenvalues of this operator
\ben
&&\Lambda^2=\hbar^2\left[j(j+1)-l(l+1)+{1\over 4}\right]^2+(mca)^2-\left({e^2\over
c}\right)^2,\\
&&j=l\pm 1/2,\ \ \ \ \ l=0,1,2,\ldots\, .
\een
Further, if $j=l+1/2$, then
\ben
\Lambda^2=\hbar^2(l+1)^2+(mca)^2-(e^2/c)^2,
\een
and if  $j=l-1/2$, then
\ben
\Lambda^2=\hbar^2l^2+(mca)^2-(e^2/c)^2.
\een
Now it is not difficult to find the eigenvalues of the operator $\hat\Lambda$ (\ref{lambd2}):
\ben
\Lambda=-\hbar\sqrt{(l+1)^2+(mca/\hbar)^2-(e^2/\hbar c)^2},
\een
for  $j=l+1/2$ and
\ben
\Lambda=\hbar\sqrt{l^2+(mca/\hbar)^2-(e^2/\hbar c)^2},
\een
for $j=l-1/2$. Finally, we can easily find the eigenvalues of the operator
$\hat \Lambda(\hat\Lambda+\hbar)$:
\ben
\Lambda(\Lambda+\hbar)=\hbar^2 l^{*}(l^{*}+1),
\een
where the quantum number
\be
l^{*}=\sqrt{(j+1/2)^2+(mca/\hbar)^2-(e^2/\hbar
c)^2}-{1/2}\mp{1/2}.
\label{kvantch}
\ee
In Eq.~(\ref{kvantch}), the upper sign is for $j=l+1/2$ while the lower sign is for $j=l-1/2$.

Now, substituting the operator $\hat\Lambda(\hat \Lambda+\hbar)$
with its eigenvalues in Eq.~(\ref{sfer}), we can write down the equation for the radial part $R$
of $\bar \psi$:
\be
\left\{-{\hbar^2\over 2m}{1\over r}{d^2\over dr^2}r+{\hbar^2l^{*}(l^{*}+1)\over 2mr^2}
-\left({E\over mc^2}e^2-mc^2a\right){1\over r}\right\}\, R={E^2-m^2c^4\over
2mc^2}R.
\label{rad}
\ee

\section{Energy eigenvalues. Discussion of the results}
\setcounter{equation}{0}

Formally, equation (\ref{rad}) coincides with the non-relativistic Schr\"odinger equation for the Kepler
problem with the energy
\ben
E^{*}={E^2-m^2c^4\over 2mc^2}
\een
and the charge squared
\be
e^{*2}={E\over mc^2}e^2-mc^2a.
\label{kvzar}
\ee

For the existence of the bound states, it is necessary that the ``potential energy" in Eq.~(\ref{rad})
should be of the attractive nature or, otherwise said, the charge squared $e^{*2}$
should be a positive value $e^{*2}>0$.

Then one can write the Bohr formula for the energy levels $E^{*}$:
\ben
E^{*}=-{me^{*4}\over 2\hbar^2(n_r+l^{*}+1)^2},
\een
where $n_r=0,1,2,\ldots$ is the radial quantum number, hence, the equation for $E$
\ben
{E^2-m^2c^4\over 2mc^2}=-{m(Ee^2/mc^2-mc^2a)^2\over 2\hbar^2
n^{*2}},
\een
where we introduced the ``principal'' quantum number
$$
n^{*}=n_r+l^{*}+1=n_r+\sqrt{(j+1/2)^2+(a/\lambda)^2-\alpha^2}+{1/2}\mp{1/2},
$$
where $\alpha=e^2/\hbar c$ is the fine structure constant and $\lambda=\hbar/mc$ is the Compton length.

Solving this equation we finally find for the energy spectrum
\be
E={mc^2\over 1+(\alpha/n^{*})^2}\left[{a\over \lambda}{\alpha\over n^{*2}}+
\sqrt{1+{\alpha^2-(a/\lambda)^2\over n^{*2}}}\right].
\label{enspektr}
\ee

The wave functions $R$ are the usual radial functions of the non-relativistic
hydrogen problem with the quantum numbers $n^{*}$,
$l^{*}$ and with the charge squared $e^{*2}$. In order to determine the full wave function $\psi$
it is  necessary to substitute the function $\bar \psi$ into Eq.~(\ref{2.4}) in the form
of the product of the radial function $R$ and the spherical spinor.

The bound states exist when $e^{*2}>0$. Using Eq.~(\ref{kvzar}) for the energy $E$
in Eq.~(\ref{enspektr}) we find the condition for the parameter $a$ that does not depend
on the quantum number $n^{*}$: $a<{e^2/ mc^2}.$ Thus the bound states exist for arbitrary
negative values of the parameter $a$. If the length parameter $a>0$, it should be
smaller than the electron classical radius. In other words, these are the distances $1/\alpha \backsimeq 137$-fold
smaller than the Compton  length $\lambda$, where the well-defined notion of the particle coordinate is lost.
It is interesting to note that for $a=e^2/mc^2$ the energy $E=mc^2$, hence, there is just one level.

On the other hand, if the parameter $a=0$, then from Eq.~(\ref{enspektr})
we get the well-known formula for a fine structure of the hydrogen atom energy spectrum, for which there is
the exact solution of the Dirac equation.

In the absence of the Coulomb interaction ($e^2=0$) the bound states exist for $a<0$ with the energy levels
$$E=mc^2\sqrt{1-\left({a\over \lambda n^{*}}\right)^2},$$
where the quantum number $l^{*}$, contained in $n^{*}$, is determined by equation (\ref{kvantch}) for $e^2=0$.

For the ground state, for $n^{*}=\sqrt{1+(a/\lambda)^2-\alpha^2}$, one obtains from Eq.~(\ref{enspektr})
\ben
E={mc^2\over 1+(a/\lambda)^2}\left[{a\over
\lambda}\alpha+\sqrt{1+(a/\lambda)^2-\alpha^2}\right].
\een
For $\alpha=0$  the ground state energy ($a<0$)
$$E={mc^2\over \sqrt{1+(a/\lambda)^2}}.$$
This quantity plays the role of the particle rest energy, thus, the mean value of the effective
mass, i.e.,
$$m^{*}={m\over\sqrt{1+(a/\lambda)^2}}$$
is smaller than the mass $m$, which corresponds to the negative parameter $a$. If parameter
$|a| \gg \lambda$, then the ground state energy $E=\hbar
c/|a|$ is of the nature of the Casimir energy concentrated in the volume $\sim|a|^3$.

Let us now consider the non-relativistic limit $c\to \infty$. We assume
that the natural length scale for the quantity $a$ is the classical radius
of the electron, which, in our problem, is the upper limit for $a$; hence
we put $a = \bar a e^2/mc^2$, $\bar a < 1$. We believe that the numeric value of $\bar
a$ does not depend on any fundamental constants, and that,
ultimatelly, this value is the initial characteristic of the particle.
Now the formula for the energy takes the form
\be
\label{e}
&&E={mc^2\over 1+(\alpha/n^{*})^2}\left[\left({\alpha\over n^{*}}\right)^2\bar a+\sqrt{1+\left({\alpha\over n^{*}}\right)^2
(1-\bar a^2)}\right],\\
&&n^{*}=n_r+\sqrt{\left(j+{1\over 2}\right)^2+\alpha^2(\bar
a^2-1)}+{1\over 2}\mp {1\over 2}.
\nonumber
\ee

We expand this expression in power series over $\alpha$ up to $\alpha^2$:
\be
\label{4.4}
&&E=mc^2-{me^4\over 2\hbar^2 n^2}(1-\bar a)^2\\ \nonumber
&&-{me^4\over 2\hbar^2 n^4} \alpha^2 (1-\bar a)^3
\left[{n\over j+1/2}(1+\bar a)-{3\over 4}(1+\bar a/3)\right],
\ee
where $n=n_r+l+1$ is the principal quantum number.  For $\bar a=0$ we have the familiar fine structure formula.
As we can see, the dependence of the mass upon the coordinate also deforms the non-relativistic
term (i.e., the Bohr formula) in the way as if the particles mass had been substituted with $m(1-\bar a)^2$.
For $\bar a>0$ this effect has the same sign as the correction accounting for the finiteness of the nucleus mass.
The degeneracy of terms, in particular of $S_{1/2}$ and $P_{1/2}$, holds because we have the specific
dependence of the mass on $r$. If, in expansion (1.1), we also left the next terms, this degeneracy would be removed.

\section{Conclusion}

We have found the exact solution of the Dirac equation for a particle with the position-dependent
mass, which might be usefull in the study of the corresponding non-relativistic problem as
a reference result. The next terms, which we have neglected in this work (in particular the dipolar one)
and which are responsible for the super-fine structure of the energy spectrum, can be taken into account
by means of the standard perturbation theory.

Let us return to the ordering problem in the kinetic energy operator in
the Schr\"odinger equation when the mass of the particle depends on the coordinate.
We have already seen that the corrections from this dependence for the non-relativistic limit
are due to the fact that in the mass this dependence occurs to be $\sim 1/c^2$.
The mechanism of the apperance  of the abovesaid correction is quite simple.
The particle rest energy  is $m^{*}c^2=mc^2(1+\bar a e^2/mc^2 r)=mc^2+\bar a e^2/r$.
In the linear approximation, according to perturbations theory, this energy
is $(mc^2+\bar a\la e^2/r\ra )=mc^2+\bar a e^2/a_{\rm B}n^2=mc^2+\bar a me^4/\hbar^2 n^2$,
where $a_{\rm B}=\hbar^2/me^2$ is the Bohr radius.
This expression, together with the zeroth approximation for $E$, in accordance with the Bohr formula,
yields $mc^2-me^4(1-2\bar a )/2\hbar^2 n^2$, which,
in the linear approximation in $\bar a$ coincides with formula (\ref{4.4}).
Otherwise said, only on condition that the dependence of the mass upon the coordinate is $\sim
1/c^2$, we deform the non-relativistic expression for the energy levels. At the same time, it is
irrelevant how we order the momentum operator with the inverse mass in the
kinetic energy operator because the corrections are of the order of $1/c^2$.

Thus, the issue of the form of the kinetic energy in the Schr\"odinger
equation for the position-dependent mass has, in the general case, quite a limited sense.
If this dependence arises in the non-relativistic case, as a result of the reduction of
the many-particle problem to a single-particle one, or as problems in the curved space,
the specific ordering of the operators appears rather naturally in each problem.
As a consequence, the energy levels, which depend on some ordering
parameter, differ from problem to problem. This has been shown in
Ref.~\cite{Tkachuk}, where the Schr\"odinger equation with the Coulomb  potential is solved
with some other dependence of the mass on the radial variable.
Yet it is important that for highly excited states, i.e., for large values of quantum numbers,
the dependence of the energy on the ordering ``parameter'' is, probably, insignificant, and the energy
can be found using the quasi-classical Bohr--Sommerfeld quantization method.
For instance, for the dependence of $m^{*}$  on  $r$ given by Eq.~(\ref{2.2})
the direct solution of the Schr\"odinger equation, in which $1/\sqrt{m^{*}}$ stands leftwards
and rightwards of the momentum square, obviously yields the same energy levels as the Bohr--Sommerfeld method.
Finally, there is also another argument in support of our theory,
which follows from the results of Ref.~\cite{Tkachuk}, that for large quantum numbers the
energy does not depend on the inverse mass and momentum ordering ``parameter''.

The author is very grateful to Volodymyr Tkachuk for insightful discussions on this problem.

\end{document}